\documentclass[USenglish,twocolumn]{article}

\usepackage[a4paper, margin=2cm]{geometry}
\usepackage{amsmath,amssymb,graphicx}
\usepackage{layouts}
\usepackage{xcolor}
\usepackage{lineno}
\usepackage{lipsum}
\usepackage{tocloft}
\usepackage{hyperref}
\usepackage{authblk}

\newcommand{\G}{\mathrm{G}}
\renewcommand{\d}{\mathrm{d}}

\title{Space-Time Graded-Index Interfaces and Related Chirping}
\author[1]{Zhiyu Li\thanks{lizhiyu@stu.xjtu.edu.cn}}
\author[1]{Xikui Ma}
\author[2]{Klaas De Kinder} 
\author[2]{Amir Bahrami}
\author[2]{Christophe Caloz}

\affil[1]{Xi’an Jiaotong University, Xi’an 710049, China}
\affil[2]{KU Leuven, Leuven 3001, Belgium}

\date{}

\begin{document}

\maketitle

\begin{abstract}
Space-time modulated systems have recently emerged as a powerful platform for dynamic electromagnetic processing in both space and time. Most of the related research so far has assumed abrupt parameter profiles. This paper extends the field to generalized graded-index (GRIN) interfaces, which are both more practical than ideal profiles and offer new avenues for wave manipulations. It presents an exact solution for wave propagation across arbitrary space-time modulated GRIN interfaces and describes versatile chirping effects. The solution is based on a generalization of the impulse response method from linear time-invariant to linear \emph{space-time-varying} systems. The proposed framework shows that space-time GRIN systems represent a novel approach for generating a new form of chirping that is not inherently based on dispersion, with promising applications in pulse shaping and signal processing.
\end{abstract}




\section{Introduction} \label{sec:intro}

Space-time modulation systems---media whose parameters vary dynamically in both space and time under the influence of an external traveling-wave modulation---have recently enabled a range of novel applications, including magnetfree non-reciprocity~\cite{Estep_2014_nonreciprocity}, frequency transitioning~\cite{Taravati_2018_aperiodic}, parametric amplification~\cite{Tien_1958_parametric,Pendry_2021_gain} and the breaking of fundamental bounds~\cite{Shlivinski_2018_BF}.

\begin{figure}[ht!]
\centering
\includegraphics[width=\columnwidth]{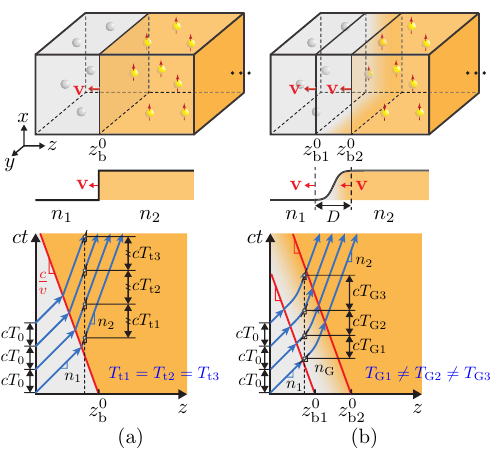}
\caption{%
Space-time modulation interfaces, with basic structures (top) and space-time diagrams in the harmonic-wave regime (bottom). (a)~Step-index modulation, where the refractive index changes abruptly from $n_1$ to $n_2$ at an interface initially located at $z_{\mathrm{b}}^{0}$ and moving with a constant velocity $\mathbf{v}$. (b)~Graded-index (GRIN) modulation, where the refractive index traditions smoothly from $n_1$ to $n_2$ via $n_\G$ over the interval $[z_{\mathrm{b}1}^{0},z_{\mathrm{b}2}^{0}]$ with width $D=z_{\mathrm{b}2}^{0}-z_{\mathrm{b}1}^{0}$.
}
\label{fig:Illustration}
\end{figure}

Space-time interfaces are the the fundamental building blocks---or `meta-atoms'--of spacetime modulation systems. Their understanding and control are therefore essential. These interfaces can be of two distinct types, as illustrated in Fig.~\ref{fig:Illustration}. The first is the \emph{step-index interfaces}, where the transition width is much smaller than the wavelength, as shown in Fig.~\ref{fig:Illustration}(a). Such interfaces have been extensively studied in the literature~\cite{Biancalana_2007_dynamics,Deck_2018_wave} and are known to induce a uniform frequency shifting of the incoming wave. The second type is the \emph{graded-index (GRIN) interfaces}. These interfaces, shown in Fig.~\ref{fig:Illustration}(b), may be considered more practical, as physical modulations necessarily occur gradually at the microscopic level~\cite{Gaafar_2019_front,Sloan_2022_twophoton,Ball_2025_knife}, and can also support a broader range of wave transformations. In particular, they induce nonuniform frequency transitions and produce wave chirping. While some studies have addressed the limiting case of pure-time GRIN interfaces~\cite{Felsen_1970_wave,Fante_1971_transmission,Hadad_2020_soft,Hayran_2022_energy}, general space-time GRIN interfaces~\cite{Deck_2023_yeecell} remain an essentially unexplored topic.

This paper investigates the wave transformations induced by space-time GRIN interfaces, introducing a generalization of the impulse response method from linear time-invariant (LTI) to linear (space-)time-varying (LTV) systems. Unlike the Wentzel–Kramers–Brillouin (WKB)~\cite{Felsen_1970_wave,Fante_1971_transmission,Hadad_2020_soft} or transfer-matrix method (TMM)~\cite{Deck_2023_yeecell} commonly used in prior works, this approach provides an \emph{exact} solution. We derive closed-form solutions for both the analysis and synthesis problems, and describe the corresponding chirping effects. All the results are validated through full-wave simulations.

For simplicity, we restrict our attention to systems that are 1+1D, with dimensions $z$ and $t$, involving GRINs of uniform velocity, i.e., $\mathbf{v}=v\mathbf{\hat{z}}=\text{const.}$, boundary and intrinsic impedance matching, i.e., $\eta_m=\sqrt{\mu_m(z,t)/\epsilon_m(z,t)}\neq\eta_m(z,t)=\text{const.}$ for $m=1,\G,2$ and no dispersion, i.e., $n_m\neq{n_m}(\omega)$ for $m=1,\G,2$.

\section{Generalized Impulse Response Method} \label{sec:method}

\begin{figure}[ht!]
\centering
\includegraphics[width=\columnwidth]{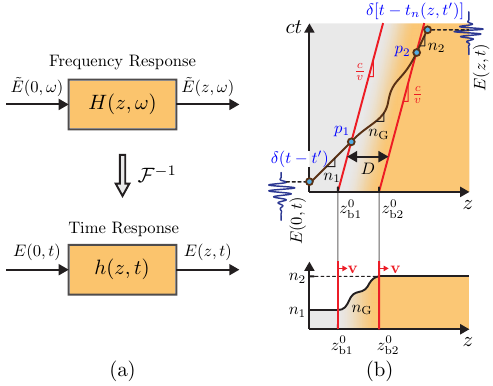}
\caption{Generalized impulse response method applied to the space-time GRIN interface system in Fig.~\ref{fig:Illustration}(b). (a)~Responses in the frequency and time domains. (b)~Response of an arbitrary space-time GRIN modulation system. The top panel shows the space-time diagram with the trajectory of the impulse $\delta(t-t')$ and its output $\delta[t-t_{n}(z,t')]$, where $p_{1,2}$ are the intersection points of the impulse with the two boundaries of the GRIN layer, initially located at $z_{\mathrm{b}1,2}^0$. The bottom panel shows the initial refractive index profile $n(z,t=0)$, which moves at a uniform velocity $\mathbf{v}$.}
\label{fig:Impulse}
\end{figure}

In this section, we shall derive the generalized impulse for the GRIN system in Fig.~\ref{fig:Illustration}(b), which is generically represented in Fig.~\ref{fig:Impulse}.

In the frequency domain, the one-dimensional Helmholtz equation may be expressed as
\begin{equation} \label{eq:1D_Wave_Equation}
    \frac{\partial^2 \tilde{E}}{\partial z^2}+k^2(\omega) \tilde{E}=0,
\end{equation}
where $k(\omega)=\omega\sqrt{\mu\epsilon}=\omega n/c$ is the wavenumber.
The general solution to Eq.~\eqref{eq:1D_Wave_Equation} can be written as
\begin{subequations}\label{eq:frequency_response}
    \begin{equation}
        \tilde{E}(z,\omega)=H(z,\omega)\tilde{E}(0,\omega),
    \end{equation}
    where
    \begin{equation}
        H(z,\omega)=\mathrm{e}^{\mathrm{i}k(\omega)z}
    \end{equation}
    represents the frequency response of the system.
\end{subequations}

Applying the temporal inverse Fourier transformation to Eqs.~\eqref{eq:frequency_response}, as suggested in Fig.~\ref{fig:Impulse}(a), we obtain the corresponding time-domain relation
\begin{equation} \label{eq:convolution}
        E(z,t)=\int^{\infty}_{-\infty} h(z,t,t')E(0,t')~\mathrm{d}t',
\end{equation}
where $h(z,t,t')$ is the impulse response~\cite{Orfanidis_2002_electromagnetic}, representing the system’s response to the impulse $E(0,t)=\delta(t-t')$\footnote{The impulse response acts as a time-domain version of the Green's function in the space-domain relation $\mathbf{E}(\mathbf{r})=\mathrm{i}\omega\mu\int_{V}G(\mathrm{r},\mathbf{r}')\mathbf{J}(\mathbf{r}')~\mathrm{d}V'$~\cite{Novotny_2012_principles}, with $E(0,t')$ in Eq.~\eqref{eq:convolution} playing the role of the current source $\mathbf{J}(\mathbf{r}')$ and $h(z,t,t')$ corresponding to the Green's function $G(\mathbf{r},\mathbf{r}')$.}.

In LTI systems, the impulse response depends solely on the time difference between the output and input signals, following the time-shift invariance property $h(z,t,t')=h(z,t-t')$~\cite{Saleh_2019_fundamentals}. Xiao et al. extended the impulse response method from LTI to purely time-varying systems~\cite{Xiao_2011_spectral}, where the lack of time invariance--or non-stationarity--leads to an impulse response that independently depends on $t$ and $t'$, i.e.,
\begin{equation}
    h(z,t,t')\neq h(z,t-t').
\end{equation}

In space-time varying systems, $h(z,t,t')$ also involves the modulation velocity $v$, which further increases the complexity of the analysis. We now introduce the generalization of the impulse response method to space-time GRIN systems, explicitly incorporating $v$, with the aid of Fig.~\ref{fig:Impulse}(b), where a space-time diagram illustrates the impulse trajectory. 

As shown in the top panel of Fig.~\ref{fig:Impulse}(b), the input impulse experiences a propagation delay as it propagates through the system, resulting in the impulse response
\begin{equation} \label{eq:ST_IR}
    h(z,t,t')=\delta[t-t_{n}(z,t')],
\end{equation}
where $t_{n}(z,t')$ is the arrival time of the impulse at the position $z$ assuming an input time of $t'$ in the GRIN system with space-time varying refractive index $n$ [black line in the top panel of Fig.~\ref{fig:Impulse}(b)]\footnote{In a non-dispersive system, the impulse response exhibits the form of a delta function as all the frequency components propagate at the same velocity, and the input impulse remains undistorted. In contrast, in dispersive systems, not considered here, the frequency-dependent group velocity causes temporal spreading, and the impulse response becomes a more complex function of $t'$~\cite{Orfanidis_2002_electromagnetic}.}. The function $t_{n}(z,t')$ is determined by the wave trajectory equation
\begin{equation} \label{eq:ODE}
    c\frac{\mathrm{d}t_{n}}{\mathrm{d}z}=n(z,t_{n}),
\end{equation}
subject to the boundary conditions
\begin{equation} \label{eq:ODE_conditiond}
    t_{n}(0)=t' \quad\text{and}\quad t_{n}(z_{\mathrm{b}1,2})=t_{\mathrm{b}1,2},
\end{equation}
where $z_{\mathrm{b}1,2}$ and $t_{\mathrm{b}1,2}$ are the coordinates of the intersection points $p_{1,2}$ of the impulse with the first and second modulation interfaces, respectively [Fig.~\ref{fig:Impulse}(b)]. 

To derive the electromagnetic field in each region of the space-time GRIN system [Fig.\ref{fig:Impulse}(b)], we will apply the following four steps to each of the three regions in Fig.~\ref{fig:Impulse}(b):
i)~determine the trajectory equation and corresponding boundary condition for the concerned region;
ii)~solve the trajectory equation for the arrival time $t_n(z,t')$;
iii)~substitute the resulting $t_n(z)$ into Eq.~\eqref{eq:ST_IR} to obtain the impulse response $h(z,t,t')$;
iv)~substitute that response into Eq.~\eqref{eq:convolution} to evaluate the output field $E(z,t)$.

\section{Field Solutions} \label{sec:derivations}

In this section, we shall derive the general field solutions for the GRIN system in Fig.~\ref{fig:Illustration}(b) using the generalized impulse response derived in Sec.~\ref{sec:method}.

We now proceed to derive the field expressions in a general GRIN system, as illustrated in Fig.~\ref{fig:Impulse}(b), following the four-step procedure outlined in Sec.~\ref{sec:method}. The refractive index profile along the impulse trajectory is given by
\begin{equation} \label{eq:n}
    n(z,t)= \begin{cases} n_1, & 0<z<z_{\mathrm{b}1}^{0}+v t, \\ n_{\mathrm{G}}(z-vt), & z_{\mathrm{b}1}^{0}+v t<z<z_{\mathrm{b}2}^{0}+v t, \\ n_2, & z>z_{\mathrm{b}2}^{0}+v t, \end{cases}
\end{equation}
where $n_1$ and $n_2$ are the (constant) refractive indices of media~1 and~2, respectively, and 
\begin{equation} \label{eq:zb120}
    z_{\mathrm{b}2}^{0}=z_{\mathrm{b}1}^{0}+D.
\end{equation}
The system can then be divided into the three corresponding regions, which we will address one by one.

\subsection{First-Medium ($n_1$) Region} \label{sec:E1}
Substituting $n(z,t_n)=n_1$ into Eq.~\eqref{eq:ODE}, we obtain the wave trajectory equation
\begin{subequations} \label{eq:ODE_n1}
    \begin{equation}
    c\frac{\mathrm{d}t_n}{\mathrm{d}z}=n_1,
\end{equation}
where the boundary condition is
\begin{equation} \label{eq:initial}
    t_n(0)=t'.
\end{equation}
\end{subequations}
Solving Eq.~\eqref{eq:ODE_n1} for $t_n(z,t')$ yields then the impulse trajectory function
\begin{equation} \label{eq:to_m1}
    t_n(z,t')=t'+\frac{n_1}{c}z,
\end{equation}
whose insertion into Eq.~\eqref{eq:ST_IR} provides the impulse response
\begin{equation} \label{eq:ip_m1}
    h(z,t,t')=\delta\left(t-t'-n_1\frac{z}{c}\right).
\end{equation}
Note that the argument of the impulse function in this relation represents a \emph{traveling-wave}, due to the uniform nature of the propagation medium. Finally, substituting Eq.~\eqref{eq:ip_m1} into Eq.~\eqref{eq:convolution} and solving for the output field $E(z,t)$, we get the field,
\begin{equation} \label{eq:E1}
    E_{1}(z,t)=E\left(0,t-n_1\frac{z}{c}\right),
\end{equation}
where $E(\cdot,\cdot)$ is an arbitrary field function (e.g., harmonic plane wave or Gaussian pulse) of the space (first entry) and time (second entry) variables\footnote{In Eq.~\eqref{eq:E1}, the bracket $(\cdot)$ indicates a functional argument. At other places in the paper, the argument may involve the square bracket $[\cdot]$ or the curl bracket $\{\cdot\}$, because we use the bracket precedence order $(\cdot)-[\cdot]-\{\cdot\}$. Whether the brackets indicate an argument or a multiplicative factor should be clear everywhere from the context.}.

\subsection{GRIN ($n_\G$) Region} \label{sec:EG}
Substituting now $n(z,t_n)=n_{\mathrm{G}}(z-vt_n)$ into Eq.~\eqref{eq:ODE}, we obtain the wave trajectory equation in the GRIN region,
\begin{equation} \label{eq:ODE_nG}
    c\frac{\mathrm{d}t_{n}}{\mathrm{d}z}=n_{\mathrm{G}}(z-vt_{n}).
\end{equation}
The corresponding space-time boundary condition with medium~1 corresponds to the intersection point $p_1$ [see Fig.~\ref{fig:Impulse}(b)], whose coordinates are related as
\begin{subequations} \label{eq:P1}
\begin{equation}
    t_{\mathrm{b}1}=t'+\frac{n_1}{c}z_{\mathrm{b}1}
\end{equation}
and
\begin{equation}
 z_{\mathrm{b}1}=z_{\mathrm{b}1}^{0}+v t_{\mathrm{b}1}.
\end{equation}
\end{subequations}
Solving Eqs.~\eqref{eq:P1} for $z_{\mathrm{b}1}$ and $t_{\mathrm{b}1}$, we find the coordinates of $p_1$ to be
\begin{equation} \label{eq:zb1_Tb1}
     z_{\mathrm{b}1}=\frac{v t'+z_{\mathrm{b}1}^{0}}{1-n_1 v/c} \quad\text{and}\quad  t_{\mathrm{b}1}=\frac{t'+n_1 z_{\mathrm{b}1}^{0}/c}{1-n_1 v/c}.
\end{equation}

To solve Eq.~\eqref{eq:ODE_nG} with the related boundary condition, we let $\xi=z-v t_n$, which implies $\frac{\mathrm{d}\xi}{\mathrm{d}z}=1-v\frac{\mathrm{d}t_n}{\mathrm{d}z}$, or
\begin{equation} \label{eq:dt_dxi}
    \frac{\mathrm{d}t_n}{\mathrm{d}z}=\frac{1}{v}\left(1-\frac{\mathrm{d}\xi}{\mathrm{d}z}\right).
\end{equation}
Substituting now Eq.~\eqref{eq:dt_dxi} into Eq.~\eqref{eq:ODE_nG}, and separating the $\xi$ and $z$ terms, leads to the differential equation
\begin{equation} \label{eq:ODE_xi}
    \frac{\mathrm{d}\xi}{1-n_{\mathrm{G}}(\xi)v/c}=\mathrm{d}z,
\end{equation}
which integrates to
\begin{equation} \label{eq:int_ng}
    \int \frac{1}{1-n_{\mathrm{G}}(\xi)v/c}~\mathrm{d}\xi=z+C_{\G},
\end{equation}
where $C_{\G}$ is an integration constant.
For generalization to arbitrary GRIN profiles, we define the left-hand side term of this relation as the function
\begin{equation} \label{eq:F}
    F(\xi)=\int \frac{1}{1-n_{\mathrm{G}}(\xi)v/c}~\mathrm{d}\xi.
\end{equation}
which allows to express Eq.~\eqref{eq:int_ng} in the compact form
\begin{equation} \label{eq:F_xi}
    F(\xi)=z+C_{\G}.
\end{equation}
To determine the integration constant $C_{\G}$ in this relation, we first apply the boundary condition $t_n(z_{\mathrm{b}1})=t_{\mathrm{b}1}$ (point $p_1$), which yields
\begin{equation} \label{eq:F_zb1_Tb1}
    F(z_{\mathrm{b}1}-vt_{\mathrm{b}1})=z_{\mathrm{b}1}+C_{\G}.
\end{equation}
We next substitute Eqs.~\eqref{eq:zb1_Tb1} into this relation and solve for $C_{\G}$, we obtain
\begin{equation} \label{eq:C_tp}
    C_{\G}(t')=-\frac{v}{1-n_1 v/c}t'-\frac{z_{\mathrm{b}1}^{0}}{1-n_1 v/c}+F(z_{\mathrm{b}1}^{0}).
\end{equation}
Finally, substituting Eq.~\eqref{eq:C_tp} and the relation $\xi=z-v t_n$ into Eq.~\eqref{eq:F_xi}, and solving for $t_n$, we find
\begin{equation} \label{eq:tz_GRIN}
    t_n(z,t')=-\frac{1}{v}F^{-1}[z+C_{\G}(t')]+\frac{z}{v},
\end{equation}
where $F^{-1}(\cdot)$ is the inverse function of $F(\cdot)$.
    
Substituting Eq.~\eqref{eq:tz_GRIN} into Eq.~\eqref{eq:ST_IR}, we obtain now the impulse response
\begin{equation} \label{eq:ip_mG}
    h(z,t,t')=\delta\left\{t+\frac{1}{v}F^{-1}[z+C_{\G}(t')]-\frac{z}{v}\right\},
\end{equation}
where $C_{\G}(t')$ is given in Eq.~\eqref{eq:C_tp} and $F(\cdot)$ is defined in Eq.~\eqref{eq:F}.

Finally, substituting Eq.~\eqref{eq:ip_mG} into Eq.~\eqref{eq:convolution}, and solving for the output field $E(z,t)$, yields (see Appendix~\ref{app:E_G})
 \begin{equation} \label{eq:EG}
 \begin{aligned}
        &E_{\mathrm{G}}(z,t)=\left|\frac{1-n_1 v/c}{1-n_{\mathrm{G}}(z-vt)v/c}\right|\\& 
       \begin{aligned}
       E\Bigg[0,
       -\frac{1-n_1 v/c}{v}\int_{z_{\mathrm{b}1}^{0}+v t}^{z} \frac{1}{1-n_{\mathrm{G}}(z'-vt)v/c}~\mathrm{d}z'&\\
           -n_1\frac{z}{c}+\frac{z-z_{\mathrm{b}1}^{0}}{v}\Bigg],
       \end{aligned}
\end{aligned}
\end{equation}
where $E(\cdot,\cdot)$ is the same field function of space and time as in Eq.~\eqref{eq:E1}, since under the assumption of impedance matching the field waveform does not change across system---only its magnitude and phase change.

\subsection{Second-Medium ($n_2$) Region} \label{sec:E2}
Substituting now $n(z,t_n)=n_2$ into Eq.~\eqref{eq:ODE}, we obtain
\begin{equation} \label{eq:ODE_n2}
    c\frac{\mathrm{d}t_n}{\mathrm{d}z}=n_2,
\end{equation}
with space-time boundary condition with the GRIN medium corresponding to the intersection point $p_2$, whose coordinates are related as
\begin{subequations} \label{eq:P2}
\begin{equation}
    t_{\mathrm{b}2}=-\frac{1}{v}F^{-1}[z_{\mathrm{b}2}+C_{\G}(t')]+\frac{z_{\mathrm{b}2}}{v}
\end{equation}
and
\begin{equation}
z_{\mathrm{b}2}=z_{\mathrm{b}2}^{0}+v t_{\mathrm{b}2},
\end{equation}
\end{subequations}
and found by solving Eqs.~\eqref{eq:P2} for $z_{\mathrm{b}2}$ and $t_{\mathrm{b}2}$ as
\begin{subequations} \label{eq:zb2_Tb2}
    \begin{equation} 
     z_{\mathrm{b}2}=F(z_{\mathrm{b}2}^{0})-C_{\G}(t')
\end{equation}
and
\begin{equation} 
     t_{\mathrm{b}2}=\frac{F(z_{\mathrm{b}2}^{0})-C_{\G}(t')-z_{\mathrm{b}2}^{0}}{v}.
\end{equation}
\end{subequations}

Solving next Eq.~\eqref{eq:ODE_n2} for $t_n(z,t')$, we obtain the impulse trajectory function
\begin{subequations} \label{eq:tz_M2}
    \begin{equation} \label{eq:tz_M2_2}
    t_n(z,t')=\frac{n_2}{c}z+C_2,
\end{equation}
where $C_2$ is a new integration constant, which is obtained by applying the boundary condition $t_n(z_{\mathrm{b}2})=t_{\mathrm{b}2}$ [Eqs.~\eqref{eq:zb2_Tb2}, point $p_2$] to Eq.~\eqref{eq:tz_M2_2} and solving the resulting expression for $C_2$ as
\begin{equation} \label{eq:C2}
\begin{aligned}
    C_2(t')=\frac{1-n_2v/c}{1-n_1v/c}t'+\frac{1-n_2 v/c}{v}[F(z_{\mathrm{b}1}^{0}+D)&\\-F(z_{\mathrm{b}1}^{0})]
    -\frac{n_2-n_1}{1-n_1v/c}\frac{z_{\mathrm{b}1}^{0}}{c}-\frac{D}{v}&.
\end{aligned}
\end{equation}
\end{subequations}

Finally, substituting Eq.~\eqref{eq:tz_M2} into Eq.~\eqref{eq:ST_IR}, we ge the impulse response
\begin{equation} \label{eq:ip_m2}
    h(z,t,t')=\delta\left[t-\frac{n_2}{c}z-C_2(t')\right],
\end{equation}
where $C_2(t')$ is given in Eq.~\eqref{eq:C2}.

Substituting Eq.~\eqref{eq:ip_m2} into Eq.~\eqref{eq:convolution} and solving for the output field $E(z,t)$ (see Appendix~\ref{app:E_2}), we obtain the field
\begin{equation} \label{eq:E2}
\begin{aligned}
    E_{2}&(z,t)=\left|\frac{1-n_1 v/c}{1-n_2 v/c}\right| \\& 
    \begin{aligned}  
    E\Bigg\{0,\frac{1-n_1 v/c}{1-n_2 v/c} \Bigg[t-\frac{n_2}{c}z+\frac{n_2-n_1}{1-n_1 v/c}\frac{z_{\mathrm{b}1}^{0}}{c}+\frac{D}{v}&\\
    -\frac{1-n_2 v/c}{v}\int_{z_{\mathrm{b}1}^{0}+v t}^{z_{\mathrm{b}1}^{0}+D+v t}\frac{1}{1-n_{\mathrm{G}}(z'-vt)v/c}~\mathrm{d}z'\Bigg]\Bigg\}.
    \end{aligned}
\end{aligned}
\end{equation}

Equations~\eqref{eq:E1},~\eqref{eq:EG} and~\eqref{eq:E2} represent the key results of this paper. They accommodate arbitrary field waveforms $E(0,t)$ and arbitrary GRIN profiles $n_{\mathrm{G}}(z-vt)$.

\section{Chirping Physics} \label{sec:chrip}

Due to the space and time varying properties of the GRIN medium, the wave behavior in our system is fairly complex, as mathematically apparent in Eq.~\eqref{eq:EG}. In this section, we shall show that this system produces a new kind of wave chirping. 

We consider a time-harmonic incident field, which reads at $z=0$,
\begin{equation} \label{eq:Ein}
    E(0,t)=\mathrm{e}^{-\mathrm{i}\omega_{0}t},
\end{equation}
where $\omega_{0}$ is assumed to be constant. Other types of fields can be treated similarly.
Substituting Eq.~\eqref{eq:Ein} into Eq.~\eqref{eq:EG}---specifically, inserting the content of the second slot of $E(\cdot,\cdot)$ in Eq.~\eqref{eq:EG} into the relation $\phi_\G=-\omega_0t(z)$---we obtain the wave phase in the GRIN layer as
\begin{equation} \label{eq:phase}
\begin{aligned}
 \phi_{\mathrm{G}}=\omega_{0}\frac{1-n_1 v/c}{v}\int_{z_{\mathrm{b}1}^{0}+v t}^{z}\frac{1}{1-n_{\mathrm{G}}(z'-vt)v/c}~\mathrm{d}z'&\\
    +\omega_{0}\left(n_1\frac{z}{c}-\frac{z-z_{\mathrm{b}1}^{0}}{v}\right).&
\end{aligned}
\end{equation}

The related instantaneous frequency at a given position in the GRIN layer is obtained by differentiating the phase in Eq.~\eqref{eq:phase} with respect to time, yielding
\begin{equation} \label{eq:w_t}
\begin{aligned}
    \omega_{\mathrm{G}}&=-\omega_{0}\frac{1-n_1 v/c}{v}\frac{\partial }{\partial t}\left[\int_{z_{\mathrm{b}1}^{0}+v t}^{z}\frac{1}{1-n_{\mathrm{G}}(z'-vt)v/c}~\mathrm{d}z'\right]\\
    &=\omega_{0}\frac{1-n_1 v/c}{1-n_{\G}(z-vt)v/c},
\end{aligned}
\end{equation}
where we have used the Leibniz integral rule. Equation~\eqref{eq:w_t} reveals that the wave frequency in the GRIN layer varies with time, through the function $n_{\G}(z-vt)$ in the denominator, indicating a \emph{space-time chirping effect}. This effect is fundamentally different from the group-velocity dispersion (GVD) chirping effect occurring in dispersive media~\cite{Saleh_2019_fundamentals}, since no dispersion is present. It will be explained shortly.

Furthermore, the chirp parameter $\alpha$, whose sign determines whether the field is up-chirping ($\alpha>0$) or down-chirping ($\alpha<0$), is obtained by time-differentiating the instantaneous frequency in Eq.~\eqref{eq:w_t}, which gives
\begin{equation}
    \alpha=\frac{\partial \omega_{\mathrm{G}}}{\partial t}=-\omega_{0}\frac{(1-n_1 v/c)}{c[1-n_{\mathrm{G}}(z,t)v/c]^2}\frac{\partial  n_{\mathrm{G}}(z,t)}{\partial z}.
\end{equation}
This result indicates that the sign of the chirp, $\alpha\gtrless{0}$, depends on
\begin{equation} \label{eq:updown}
    (1- n_1 v/c)\partial n_{\mathrm{G}}(z,t)/\partial z\lessgtr 0.
\end{equation}
%

According to Eq.~\eqref{eq:updown}, the up- or down-chirping behavior of the system is governed by the velocity regime, which may be subluminal [$v<c/\max(n_1,n_2)$] or superluminal [$v>c/\min(n_1,n_2)$]~\cite{Caloz_2019_spacetime2}. To understand the chirping mechanism within the GRIN layer, we now focus on the subluminal case---the superluminal case can be analyzed analogously.
In the subluminal regime, where $v<c/n_1$, Eq.~\eqref{eq:updown} simplifies to
\begin{equation} \label{eq:sub_chirp}
    \partial n_{\mathrm{G}}(z,t)/\partial z\lessgtr 0,
\end{equation}
indicating that a decreasing refractive index slope leads to an increase in the instantaneous frequency, i.e., positive chirping, and vice versa, i.e., negative chirping.

\begin{figure}[ht!]
\centering
\includegraphics[width=\columnwidth]{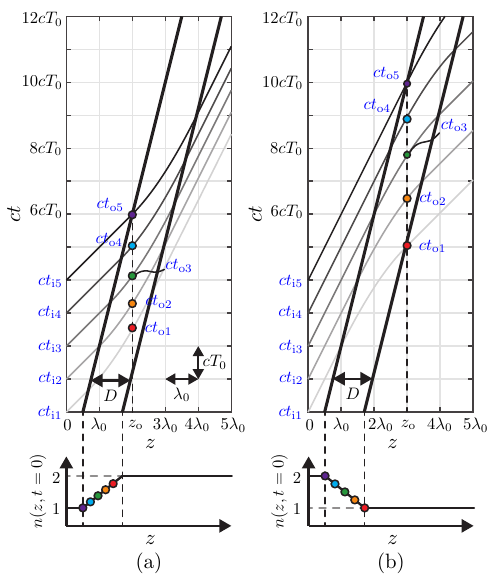}
\caption{Space-time diagrams of a linearly varying GRIN medium [Eq.~\eqref{eq:linear_nG}] in the subluminal velocity regime ($v=0.25c$) used for the chirping analysis, with (a)~a positive $n_\G$ slope, where $n_1=1, n_2=2$ and $D=1.2\lambda_{0}$ with $\lambda_{0}=cT_{0}$ being the free-space wavelength, and (b)~a negative $n_\G$ slope, where $n_1=2, n_2=1$ and $D=1.2\lambda_{0}$. The bottom panels show the corresponding initial refractive index profiles, $n(z,t=0)$.}
\label{fig:Analysis}
\end{figure}

For simplicity, but without loss of generality, we consider the simplest GRIN profile, the linear profile. 
\begin{equation} \label{eq:linear_nG}
    n_{\G}(z-vt)=n_1+\frac{n_2-n_1}{D}(z-vt-z_{\mathrm{b}1}^{0}).
\end{equation}
Figure~\ref{fig:Analysis} represents the corresponding trajectories for the wave crests incident at $z=0$ at the times $t_{\mathrm{i}1,2,3,4,5}$,
as derived from the analytical solutions in the different regions, given by Eqs.~\eqref{eq:E1},~\eqref{eq:EG} and~\eqref{eq:E2}.
Figure~\ref{fig:Analysis}(a) considers the case where the refractive index increases within the GRIN layer, i.e., $n_2>n_1$, corresponding to the positive spatial gradient $\partial n_{\G}(z,t)/\partial z>0$, which leads to a down-chirping according to the condition in Eq.~\eqref{eq:sub_chirp}.
For a fixed observation position $z_{\mathrm{o}}$ within the GRIN region, the arrival times of the five incident crests are denoted as $t_{\mathrm{o}1,2,3,4,5}$. Due to the space-time variation of $n_{\G}$, each crest experiences a different local refractive index at the observation point, as indicated by the color-coded dots in the figure.
The latest crest (purple dot) propagates fastest, as it has just entered the GRIN region and hence encounters the lowest refractive index. In contrast, the earliest crest (red dot) travels slowest, as it is near the exit of the GRIN region and hence sees the highest refractive index. As a result, the later, faster crest gradually catches up with the earlier, slower one, leading to an increasing temporal separation between adjacent crests. This corresponds to a decreasing frequency over time, i.e., down-chirping, consistent with the down-chirping condition in Eq.~\eqref{eq:sub_chirp}.
A similar mechanism applies in the case of a decreasing refractive index, i.e., $n_2<n_1$, where the crests compress in time, resulting in up-chirping [Eq.~\eqref{eq:sub_chirp}], as illustrated in Fig.~\ref{fig:Analysis}(b).

Note that the wave exiting the GRIN layer is not chirping anymore. This because at the exit points, all the crest have experience the entire GRIN profile and find therefore themselves resynchronize, now to the velocity corresponding to the second medium.

The relation~\eqref{eq:w_t} offers a practical foundation for designing systems with a prescribed chirp profile, where the frequency varies according to a desired function $f(t)$, 
\begin{equation} \label{eq:inverse}
    \omega_{\G}=f(t).
\end{equation}
To realize such a frequency evolution, one can tailor the refractive index profile of the GRIN layer, by substituting Eq.~\eqref{eq:w_t} into Eq.~\eqref{eq:inverse}, and solving for $n_{\G}$, yielding
\begin{equation} \label{eq:n_w_inver}
    n_{\mathrm{G}}=\frac{c}{v}\left[1-\frac{\omega_{0}(1-n_1 v/c)}{f(t)}\right],
\end{equation}
which provides a closed-form expression for engineering a GRIN profile generating the desired chirp.
As an example, for a linear chirp, where 
\begin{equation} \label{eq:f_linear}
    f(t)=a+b(z-v t)
\end{equation}
with $a$ and $b$ being constants, the refractive index profile becomes
\begin{equation} \label{eq:nG_linear}
    n_{\mathrm{G}}=\frac{c}{v}\left[1-\frac{\omega_{0}(1-n_1 v/c)}{a+b(z-v t)}\right].
\end{equation}

\section{Illustrative Results}

\begin{figure}[ht!]
\centering
\includegraphics[width=\columnwidth]{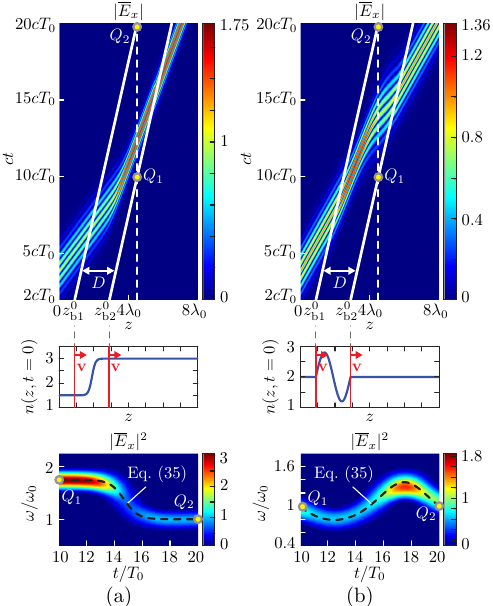}
\caption{Electric field magnitude $|E_{x}|$ across space-time GRIN interfaces, computed by Eqs.~\eqref{eq:E1},~\eqref{eq:EG} and~\eqref{eq:E2}, for the input pulse~$E(0,ct)=\mathrm{e}^{-(t-4T_0)^2/2T_0^2}\mathrm{e}^{-\mathrm{i}\omega_{0}t}$, interface velocity $v = 0.2c$, and different GRIN profiles,~(a)~a hyperbolic tangent profile, $n_{\G}(z-vt)=n_1+[(n_2-n_1)/2][1+10\tanh(z-vt-z_{\mathrm{b} 1}^0-D/2)/D]$, with $n_1=1.5$, $n_2=3$ and $D=2\lambda_0$, and (b)~a sine profile, $n_{\G}(z-vt)=n_1+[0.8+n_2-n_1]\sin[2\pi(z-vt-z_{\mathrm{b} 1}^0)/D]$, with $n_1=n_2=2$ and $D=2\lambda_0$. The top panels show the space-time diagrams of the normalized electric field magnitude [$|\overline{E}_x(z,ct)|=|E_x(z,ct)|/\max(|E_x(0,ct)|)$] under modulated Gaussian pulse excitation, where the white solid lines mark the two boundaries of the GRIN region. The middle panels show the refractive index profiles $n(z,t)$ [Eq.~\eqref{eq:n}] at $t=0$. The bottom panels show the normalized spectrograms, $||\overline{E}(z,t)|^2(t,\omega)$, with the input pulse being replaced by the quasi-continuous wave $E(0,t)=\mathrm{e}^{-\mathrm{i}\omega_{0}t}\text{rect}(t/\tau)$ (with $\tau=30 T_0$), for easier visualization, across the GRIN layer at $z=4.5\lambda_0$ between points $Q_1$ and $Q_2$. The dashed black line corresponds to the instantaneous frequency $\omega_\text{G}(t)$ given by Eq.~\eqref{eq:w_t}.}
\label{fig:FDTD}
\end{figure}

Figure~\ref{fig:FDTD} plots the electric field magnitudes across two space-time GRIN interfaces computed by Eqs.~\eqref{eq:E1},~\eqref{eq:EG} and~\eqref{eq:E2}.
Figure~\ref{fig:FDTD}(a) corresponds to a hyperbolic tangent GRIN interface profile. It may be observed that the field experiences a gradual down-chirp in the GRIN region, as expected from $\alpha<0$ [Eq.~\eqref{eq:sub_chirp}], before reaching a steady frequency in the second medium.
Figure~\ref{fig:FDTD}(b) corresponds to a sinusoidal GRIN interface profile. In this case, the field undergoes non-monotonic, twisted chirping within the GRIN layer with varying alternating chirping sign [Eq.~\eqref{eq:sub_chirp}], and eventually recovers its original frequency after exiting the modulated region.
In both cases, the closed-form field distributions in the figure have been validated against full-wave  finite-difference time-domain (FDTD) simulations~\cite{Deck_2023_yeecell,Bahrami_2023_FDTD} (see Appendix~\ref{app:FDTD}).

\begin{figure}[ht!]
\centering
\includegraphics[width=\columnwidth]{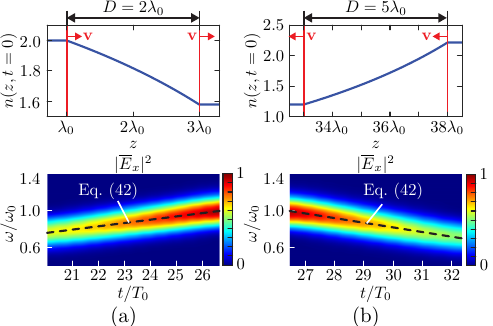}
\caption{Design of a linear-chirping [Eq.~\eqref{eq:f_linear}] GRIN system, corresponding to the refractive index profiles obtained from Eq.~\eqref{eq:nG_linear} and shown in the top panels. The input pulse is the same quasi-continuous wave as in the bottom panels of Fig.~\ref{fig:FDTD}. (a)~Up-chirping system with $n_1=2$, $n_2=1.58$, $v=0.3c$ (subluminal regime), $D=2\lambda_0$, $a=1$ and $b=-0.4$. (b)~Down-chirping system with  $n_1=1.2$, $n_2=2.22$, $v=-0.85c$ (superluminal regime), $D=5\lambda_0$, $a=1$ and $b=-0.2$. The top and bottom panels show the GRIN profiles at $t=0$ and the corresponding spectrograms, $|E(z,t)|^2(t,\omega)$, respectively. The dashed black lines represent the target linear-chirp function $f(t)$ [Eq.~\eqref{eq:f_linear}].}
\label{fig:FDTD_Inverse}
\end{figure}

Figure~\ref{fig:FDTD_Inverse} presents a linear-chirping GRIN design using~Eq.~\eqref{eq:nG_linear}.
In Fig.~\ref{fig:FDTD_Inverse}(a), the design is performed in a co-moving (up-chirping) subluminal regime with $v=0.3c$, while in Fig.~\ref{fig:FDTD_Inverse}(b), it corresponds to a contra-moving (down-chirping) superluminal regime with $v=-0.8c$.
The spectrograms in the bottom panels, corresponding to the GRIN profiles in the top panels, precisely match the theoretical predictions (see Appendix~\ref{app:FDTD}).
%

\section{Conclusion and Discussion}
We have presented an exact electromagnetic solution to the problem of wave propagation across arbitrary space-time modulated GRIN interfaces and a detailed description of the related chirping effects.
This solution offers a novel approach to chirp generation, which can be realized using switched transmission lines~\cite{Taravati_2016_Mixer,Moussa_2023_observation} and other wave modulation techniques~\cite{Caloz_GSTEMs,Gaafar_2019_front,Sloan_2022_twophoton,Ball_2025_knife}.
Our findings are significant for linear pulse shaping applications and provide a foundation for extending the study to space-time dispersive systems. Additionally, the results hold potential for future applications in dispersion compensation, paving the way for advanced control over wave propagation in dynamic media.

\appendix
\section{Derivation of Eq.~\eqref{eq:EG}} \label{app:E_G}

In this appendix, we provide a detailed derivation of Eq.~\eqref{eq:EG} given in Sec.~\ref{sec:EG}.
Substituting Eq.~\eqref{eq:ip_mG} into Eq.~\eqref{eq:convolution}, we get
\begin{equation} \label{eq:EG_conv}
\begin{aligned}
    &E_{\mathrm{G}}(z,t)=\\
    &\int^{\infty}_{-\infty} \delta\left\{t+\frac{1}{v}F^{-1}[z+C_{\G}(t')]-\frac{z}{v}\right\}E(0,t')~\mathrm{d}t'.
\end{aligned}
\end{equation}
To simplify the notation, we define the argument of the delta function in Eq.~\eqref{eq:EG_conv} as 
\begin{equation} \label{eq:g_tp}
    g_{\G}(t')=t+\frac{1}{v}F^{-1}[z+C_{\G}(t')]-\frac{z}{v}.
\end{equation}
Moreover, we use the following property of the delta function~\cite{Kusse_2006_mathematical}
\begin{equation} \label{eq:dirac_p}
    \int^{\infty}_{-\infty} 
    \delta[g_{\G}(t')]f(0,t')~\mathrm{d}t'=\frac{f(0,t'_{\G})}{|g_{\G}'(t'_{\G})|},
\end{equation}
where $g_{\G}(t'_{\G})=0$ and $g'_{\G}(\cdot)=\mathrm{d} g_{\G}(\cdot)/\mathrm{d} t'$. Therefore, we first solve for the root $t'_{\G}$ of $g_{\G}(t')=0$, i.e.,
\begin{equation} \label{eq:gtp_0}
    t+\frac{1}{v}F^{-1}[z+C_{\G}(t'_{\G})]-\frac{z}{v}=0,
\end{equation}
which gives
\begin{equation} \label{eq:C_t0p}
    C_{\G}(t'_{\G})=F(\xi)-z.
\end{equation}
Substituting Eq.~\eqref{eq:C_tp} into the left-hand side of Eq.~\eqref{eq:C_t0p} and solving for $t'_{\G}$, we find
\begin{equation} \label{eq:t0p_1}
    t'_{\G}=-\frac{1-n_1 v/c}{v}[F(\xi)-F(z_{\mathrm{b}1}^{0})]-n_1\frac{z}{c}+\frac{z-z_{\mathrm{b}1}^{0}}{v}.
\end{equation}
Applying Eq.~\eqref{eq:F} and performing the change of variables $z'=\xi'+vt$, the difference in the square brackets of Eq.~\eqref{eq:t0p_1} can be written as
\begin{equation} \label{eq:F-F}
    F(\xi)-F(z_{\mathrm{b}1}^{0})=\int_{z_{\mathrm{b}1}^{0}+v t}^{z}\frac{1}{1-n_{\mathrm{G}}(z'-vt)v/c}~\mathrm{d}z'.
\end{equation}
Substituting Eq.~\eqref{eq:F-F} into Eq.~\eqref{eq:t0p_1} yields the final expression
\begin{equation} \label{eq:t0p}
\begin{aligned}
     t'_{\G}=-\frac{1-n_1 v/c}{v}\int_{z_{\mathrm{b}1}^{0}+v t}^{z}\frac{1}{1-n_{\mathrm{G}}(z'-vt)v/c}~\mathrm{d}z'&\\
     -n_1\frac{z}{c}+\frac{z-z_{\mathrm{b}1}^{0}}{v}.&
\end{aligned}
\end{equation}

Next, we determine the derivative of $g_{\G}(t')$ [Eq.~\eqref{eq:g_tp}] with respect to $t'$ by applying the chain rule and introducing the substitution $u=z+C_{\G}(t')$, yielding
\begin{equation} \label{eq:dg_dtp}
    \frac{\mathrm{d} g_{\G}(t')}{\mathrm{d} t'}=\frac{1}{v}\frac{\mathrm{d}F^{-1}(u)}{\mathrm{d}u}\frac{\mathrm{d}C_{\G}(t')}{\mathrm{d}t'}.
\end{equation}
Let $w=F^{-1}(u)$, i.e., $u=F(w)$. Then the derivative $\mathrm{d}F^{-1}(u)/\mathrm{d}u$ on the right-hand side of Eq.~\eqref{eq:dg_dtp} can be expressed as
\begin{equation} \label{eq:dF_du}
    \frac{\mathrm{d}F^{-1}(u)}{\mathrm{d}u}=\frac{\mathrm{d}w}{\mathrm{d}u}=\frac{1}{\frac{\mathrm{d}u}{\mathrm{d}w}}=\frac{1}{F'(w)}=\frac{1}{F'[F^{-1}(u)]},
\end{equation}
where $F'(\cdot)$ denotes the derivative of $F$ with respect to its argument.
Using Eq.~\eqref{eq:C_tp}, the derivative $\mathrm{d}C_{\G}(t')/\mathrm{d}t'$ on the right-hand side of Eq.~\eqref{eq:dg_dtp} is given by
\begin{equation} \label{eq:dCtp_dtp}
    \frac{\mathrm{d}C_{\G}(t')}{\mathrm{d}t'}=-\frac{v}{1-n_1 v/c}.
\end{equation}
Substituting Eqs.~\eqref{eq:dF_du} and~\eqref{eq:dCtp_dtp} into Eq.~\eqref{eq:dg_dtp}, we obtain
\begin{equation} \label{eq:dgtp}
    \frac{\mathrm{d} g_{\G}(t')}{\mathrm{d} t'}=-\frac{1}{1-n_1 v/c}\frac{1}{F'[F^{-1}(u)]}.
\end{equation}

At the root $t'_{\G}$, using Eq.~\eqref{eq:gtp_0}, we find
\begin{equation}
    F^{-1}[u(t'_{\G})]=z-vt=\xi.
\end{equation}
Then, applying the definition in Eq.~\eqref{eq:F}, we get
\begin{equation} \label{eq:Fp_t0p}
    F'\{F^{-1}[u(t'_{\G})]\}=F'(\xi)=\frac{1}{1-n_{\mathrm{G}}(\xi)v/c}.
\end{equation}
Substituting Eq.~\eqref{eq:Fp_t0p} into Eq.~\eqref{eq:dgtp} and using the relation $\xi=z-vt$, we find the derivative of $g_{\G}(t')$ evaluated at the root
\begin{equation} \label{eq:dgt0p}
    \frac{\mathrm{d} g_{\G}(t')}{\mathrm{d} t'}\Big|_{t'_{\G}}=-\frac{1-n_{\mathrm{G}}(z-vt)v/c}{1-n_1 v/c}.
\end{equation}

Finally, substituting Eqs.~\eqref{eq:g_tp},~\eqref{eq:t0p} and~\eqref{eq:dgt0p}, along with the relation $f(0,t')=E(0,t')$, into the delta function identity [Eq.~\eqref{eq:dirac_p}],  we evaluate the integral in Eq.~\eqref{eq:EG_conv} and find the result in Eq.~\eqref{eq:EG}.

\section{Derivation of Eq.~\eqref{eq:E2}} \label{app:E_2}

In this appendix, we provide a detailed derivation of Eq.\eqref{eq:E2} given in Sec.~\ref{sec:E2}.
Substituting Eq.~\eqref{eq:ip_m2} into Eq.~\eqref{eq:convolution}, we get
\begin{equation} \label{eq:E2_conv}
    E_{2}(z,t)=\int^{\infty}_{-\infty} \delta\left[t-\frac{n_2}{c}z-C_2(t')\right]E(0,t')~\mathrm{d}t'.
\end{equation}
We define the argument of the delta function in Eq.~\eqref{eq:E2_conv} as
\begin{equation} \label{eq:g2}
    g_2(t')=t-\frac{n_2}{c}z-C_2(t'),
\end{equation}
where $C_2(t')$ is given in Eq.~\eqref{eq:C2}.

To evaluate the convolution in Eq.~\eqref{eq:E2_conv}, we use again the property~\cite{Kusse_2006_mathematical}
\begin{equation} \label{eq:dirac_2}
    \int^{\infty}_{-\infty} 
    \delta[g_{2}(t')]f(0,t')~\mathrm{d}t'=\frac{f(0,t'_2)}{|g_{2}'(t'_{2})|},
\end{equation}
where $g_{2}(t'_{2})=0$  and $g'_{2}(\cdot)=\mathrm{d} g_{2}(\cdot)/\mathrm{d} t'$.
Substituting Eq.~\eqref{eq:C2} into Eq.~\eqref{eq:E2_conv} and solving $g_2(t')=0$ for $t'_2$, we obtain
\begin{equation} \label{eq:tp2_F}
\begin{aligned}
    t'_2=\frac{1-n_1 v/c}{1-n_2 v/c} \Bigg\{t-\frac{n_2}{c}z+\frac{n_2-n_1}{1-n_1 v/c}\frac{z_{\mathrm{b}1}^{0}}{c}+\frac{D}{v}&\\
    -\frac{1-n_2 v/c}{v}[F(z_{\mathrm{b}1}^{0}+D)-F(z_{\mathrm{b}1}^{0})]\Bigg\}.&
\end{aligned}
\end{equation}
Applying Eq.~\eqref{eq:F} and performing the substitution $z'=\xi'+vt$, the difference in the square brackets
of Eq.~\eqref{eq:tp2_F} can be written as
\begin{equation} \label{eq:F_2-F}
     F(z_{\mathrm{b}1}^{0}+D)-F(z_{\mathrm{b}1}^{0})=\int_{z_{\mathrm{b}1}^{0}+v t}^{z_{\mathrm{b}1}^{0}+D+v t}\frac{1}{1-n_{\mathrm{G}}(z'-vt)v/c}~\mathrm{d}z'.
\end{equation}
Substituting Eq.~\eqref{eq:F_2-F} into Eq.~\eqref{eq:tp2_F} yields the final expression
\begin{equation} \label{eq:tp2}
\begin{aligned}
    t'_2=\frac{1-n_1 v/c}{1-n_2 v/c} \Bigg[t-\frac{n_2}{c}z+\frac{n_2-n_1}{1-n_1 v/c}\frac{z_{\mathrm{b}1}^{0}}{c}+\frac{D}{v}&\\
    -\frac{1-n_2 v/c}{v}\int_{z_{\mathrm{b}1}^{0}+v t}^{z_{\mathrm{b}1}^{0}+D+v t}\frac{1}{1-n_{\mathrm{G}}(z'-vt)v/c}~\mathrm{d}z'\Bigg].&
\end{aligned}
\end{equation}

The derivative of $g_2(t')$ [Eq.~\eqref{eq:g2}] with respect to $t'$, evaluated at the root, is then found as
\begin{equation} \label{eq:dg2_t2p}
    \frac{\d g_2(t')}{\d t'}\Big|_{t'_2}=\frac{\d g_2(t')}{\d t'}=-\frac{1-n_2 v/c}{1-n_1 v/c}.
\end{equation}

Finally, substituting Eqs.~\eqref{eq:g2},~\eqref{eq:tp2} and~\eqref{eq:dg2_t2p}, along with the relation $f(0,t')=E(0,t')$, into the delta function identity [Eq.~\eqref{eq:dirac_2}],  we evaluate the integral in Eq.~\eqref{eq:E2_conv} and find the result in Eq.~\eqref{eq:E2}.

\section{FDTD Validation} \label{app:FDTD}

\begin{figure}[ht!]
\centering
\includegraphics[width=\columnwidth]{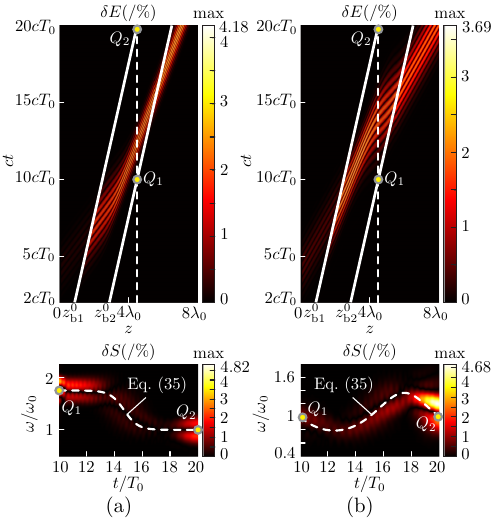}
\caption{Difference between the exact theoretical results in Fig.~\ref{fig:FDTD} and full-wave FDTD results, with difference attributed to simulation (meshing) approximation errors.}
\label{fig:FDTD_Error}
\end{figure}

Figure~\ref{fig:FDTD_Error} compares the results obtained by theory in Fig.~\ref{fig:FDTD} and by the FDTD method presented in~\cite{Deck_2023_yeecell,Bahrami_2023_FDTD}. Specifically, it plots the field (upper panels) and spectrogram (lower panels) differences
\begin{subequations}
\begin{equation} \label{eq:error}
\delta{E}=\frac{|E_{x}-E_{\mathrm{FDTD}}|}{\max(E_{x})}\times 100\%
\end{equation}
and
\begin{equation} \label{eq:error_S}
\delta{S}=\left|\frac{|E_{x}|^2-|E_{\mathrm{FDTD}}|^2}{\max(|E_{x}|^2)}\right|\times 100\%,
\end{equation}
\end{subequations}
between the theoretical and simulation results, where $E_{x}$ corresponds to the theoretical result and $E_{\mathrm{FDTD}}$ corresponds to the simulation result. The theoretical results closely match the simulation results (less than $5\%$ difference). This qualitatively validates the theory, while the fact that the theoretical results are exact (no approximation) suggests that the difference is due simulation (meshing) approximation errors, as we could verify by increasing the mesh size up to the memory capability of our computer.

\begin{figure}[ht!]
\centering
\includegraphics[width=\columnwidth]{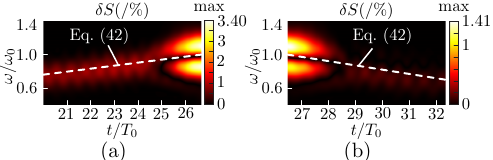}
\caption{Same as in Fig.~\ref{fig:FDTD_Error} but for Fig.~\ref{fig:FDTD_Inverse}.}
\label{fig:FDTD_Inverse_Error}
\end{figure}

Figure~\ref{fig:FDTD_Inverse_Error} shows the corresponding difference for the designed linear-chirping GRIN interfaces shown in Fig.~\ref{fig:FDTD_Inverse}, leading to the same conclusion.
%

\bibliographystyle{IEEEtran}
\bibliography{Reference.bib}

\begin{thebibliography}{10}
\providecommand{\url}[1]{#1}
\csname url@samestyle\endcsname
\providecommand{\newblock}{\relax}
\providecommand{\bibinfo}[2]{#2}
\providecommand{\BIBentrySTDinterwordspacing}{\spaceskip=0pt\relax}
\providecommand{\BIBentryALTinterwordstretchfactor}{4}
\providecommand{\BIBentryALTinterwordspacing}{\spaceskip=\fontdimen2\font plus
\BIBentryALTinterwordstretchfactor\fontdimen3\font minus \fontdimen4\font\relax}
\providecommand{\BIBforeignlanguage}[2]{{%
\expandafter\ifx\csname l@#1\endcsname\relax
\typeout{** WARNING: IEEEtran.bst: No hyphenation pattern has been}%
\typeout{** loaded for the language `#1'. Using the pattern for}%
\typeout{** the default language instead.}%
\else
\language=\csname l@#1\endcsname
\fi
#2}}
\providecommand{\BIBdecl}{\relax}
\BIBdecl

\bibitem{Estep_2014_nonreciprocity}
N.~A. Estep, D.~L. Sounas, J.~Soric, and A.~Alù, ``Magnetic-free non-reciprocity and isolation based on parametrically modulated coupled-resonator loops,'' \emph{Nat. Phys.}, vol.~10, no.~12, pp. 923--927, 2014.

\bibitem{Taravati_2018_aperiodic}
S.~Taravati, ``Aperiodic space-time modulation for pure frequency mixing,'' \emph{Phys. Rev. B}, vol.~97, no.~11, p. 115131, 2018.

\bibitem{Tien_1958_parametric}
P.~K. Tien, ``Parametric amplification and frequency mixing in propagating circuits,'' \emph{J. Appl. Phys.}, vol.~29, no.~9, pp. 1347--1357, 1958.

\bibitem{Pendry_2021_gain}
J.~B. Pendry, E.~Galiffi, and P.~A. Huidobro, ``Gain in time-dependent media—a new mechanism,'' \emph{J. Opt. Soc. Am.}, vol.~38, no.~11, pp. 3360--3366, 2021.

\bibitem{Shlivinski_2018_BF}
A.~Shlivinski and Y.~Hadad, ``Beyond the {B}ode-{F}ano bound: Wideband impedance matching for short pulses using temporal switching of transmission-line parameters,'' \emph{Phys. Rev. Lett.}, vol. 121, p. 204301, 2018.

\bibitem{Biancalana_2007_dynamics}
F.~Biancalana, A.~Amann, A.~V. Uskov, and E.~P. O'Reilly, ``Dynamics of light propagation in spatiotemporal dielectric structures,'' \emph{Phys. Rev. E}, vol.~75, p. 046607, 2007.

\bibitem{Deck_2018_wave}
Z.-L. Deck-L{\'e}ger, A.~Akbarzadeh, and C.~Caloz, ``Wave deflection and shifted refocusing in a medium modulated by a superluminal rectangular pulse,'' \emph{Phys. Rev. B}, vol.~97, no.~10, p. 104305, 2018.

\bibitem{Gaafar_2019_front}
M.~A. Gaafar, T.~Baba, M.~Eich, and A.~Y. Petrov, ``Front-induced transitions,'' \emph{Nat. Photonics}, vol.~13, no.~11, pp. 737--748, 2019.

\bibitem{Sloan_2022_twophoton}
J.~Sloan, N.~Rivera, J.~D. Joannopoulos, and M.~Solja{\v{c}}i{\'c}, ``Controlling two-photon emission from superluminal and accelerating index perturbations,'' \emph{Nat. Phys.}, vol.~18, no.~1, pp. 67--74, 2022.

\bibitem{Ball_2025_knife}
A.~Ball, R.~Secondo, D.~Fomra, J.~Wu, S.~Saha, A.~Agrawal, H.~Lezec, and N.~Kinsey, ``A space-time knife-edge in epsilon-near-zero films for ultrafast pulse characterization,'' \emph{Laser Photonics Rev.}, vol.~19, no.~5, p. 2401462, 2025.

\bibitem{Felsen_1970_wave}
L.~Felsen and G.~Whitman, ``Wave propagation in time-varying media,'' \emph{IEEE Trans. Antennas Propag.}, vol.~18, no.~2, pp. 242--253, 1970.

\bibitem{Fante_1971_transmission}
R.~Fante, ``Transmission of electromagnetic waves into time-varying media,'' \emph{IEEE Trans. Antennas Propag.}, vol.~19, no.~3, pp. 417--424, 1971.

\bibitem{Hadad_2020_soft}
Y.~Hadad and A.~Shlivinski, ``Soft temporal switching of transmission line parameters: Wave-field, energy balance, and applications,'' \emph{IEEE Trans. Antennas Propag.}, vol.~68, no.~3, pp. 1643--1654, 2020.

\bibitem{Hayran_2022_energy}
Z.~Hayran, J.~B. Khurgin, and F.~Monticone, ``$\hbar \omega$ versus $\hbar k$: dispersion and energy constraints on time-varying photonic materials and time crystals,'' \emph{Opt. Mater. Express}, vol.~12, no.~10, pp. 3904--3917, 2022.

\bibitem{Deck_2023_yeecell}
Z.-L. Deck-L\'{e}ger, A.~Bahrami, Z.~Li, and C.~Caloz, ``Generalized {FDTD} scheme for the simulation of electromagnetic scattering in moving structures,'' \emph{Opt. Express}, vol.~31, no.~14, pp. 23\,214--23\,228, 2023.

\bibitem{Orfanidis_2002_electromagnetic}
S.~J. Orfanidis, \emph{Electromagnetic {W}aves and {A}ntennas}.\hskip 1em plus 0.5em minus 0.4em\relax Rutgers University, 2002.

\bibitem{Novotny_2012_principles}
L.~Novotny and B.~Hecht, \emph{Principles of {N}ano-{O}ptics}, 2nd~ed.\hskip 1em plus 0.5em minus 0.4em\relax Cambridge university press, 2012.

\bibitem{Saleh_2019_fundamentals}
B.~E. Saleh and M.~C. Teich, \emph{Fundamentals of {P}hotonics}, 3rd~ed.\hskip 1em plus 0.5em minus 0.4em\relax John Wiley \& {S}ons, 2019.

\bibitem{Xiao_2011_spectral}
Y.~Xiao, G.~P. Agrawal, and D.~N. Maywar, ``Spectral and temporal changes of optical pulses propagating through time-varying linear media,'' \emph{Opt. Lett.}, vol.~36, no.~4, pp. 505--507, 2011.

\bibitem{Caloz_2019_spacetime2}
C.~{}Caloz and Z.-L. Deck-L{\'e}ger, ``Spacetime metamaterials—{P}art {II}: Theory and applications,'' \emph{IEEE Trans. Antennas Propag.}, vol.~68, no.~3, pp. 1583--1598, 2019.

\bibitem{Bahrami_2023_FDTD}
A.~Bahrami, Z.-L. Deck-L{\'e}ger, Z.~Li, and C.~Caloz, ``A generalized {FDTD} scheme for moving electromagnetic structures with arbitrary space-time configurations,'' \emph{IEEE Trans. Antennas Propag.}, vol.~72, no.~2, pp. 1721--1734, 2024.

\bibitem{Taravati_2016_Mixer}
S.~Taravati and C.~Caloz, ``Mixer-duplexer-antenna leaky-wave system based on periodic space-time modulation,'' \emph{IEEE Trans. Antennas Propag.}, vol.~65, no.~2, pp. 442--452, 2016.

\bibitem{Moussa_2023_observation}
H.~Moussa, G.~Xu, S.~Yin, E.~Galiffi, Y.~Ra’di, and A.~Al{\`u}, ``Observation of temporal reflection and broadband frequency translation at photonic time interfaces,'' \emph{Nat. Phys.}, vol.~19, no.~6, pp. 863--868, 2023.

\bibitem{Caloz_GSTEMs}
C.~Caloz, Z.-L. Deck-L{\'e}ger, A.~Bahrami, O.~C. Vicente, and Z.~Li, ``Generalized space-time engineered modulation ({GSTEM}) metamaterials: A global and extended perspective.'' \emph{IEEE Antennas Propag. Mag.}, vol.~65, no.~4, pp. 50--60, 2023.

\bibitem{Kusse_2006_mathematical}
B.~R. Kusse and E.~A. Westwig, \emph{Mathematical {P}hysics: {A}pplied {M}athematics for {S}cientists and {E}ngineers}, 2nd~ed.\hskip 1em plus 0.5em minus 0.4em\relax Wiley-VCH, 2006.

\end{thebibliography}

\clearpage

\end{document}